\documentclass[preprint,10pt,3p,twocolumn]{elsarticle}

\usepackage{graphicx} % Include figure files
\usepackage{epstopdf} %
\usepackage{dcolumn} % Align table columns on decimal point
\usepackage{bm} % bold math
\usepackage{ulem} %pretty font effects
\usepackage{threeparttable}

\usepackage{caption}
\usepackage{subcaption}

\usepackage{amsmath}
\usepackage{array}
\usepackage{tabulary}
\usepackage{dcolumn}% Align table columns on decimal point
\usepackage{multirow}
\usepackage{rotating}
\usepackage[usenames,dvipsnames]{color}

\makeatletter

\newcommand{\Rmnum}[1]{\expandafter\@slowromancap\romannumeral #1@}
\makeatother

\journal{International Journal of Fatigue}

\begin{document}

\begin{frontmatter}

\title{A non-linear damage accumulation fatigue model for predicting strain life at variable amplitude loadings based on constant amplitude fatigue data}

\author[isu]{P. J. Huffman}
\author[isu]{S. P. Beckman\corref{cor1}}
\ead{sbeckman@iastate.edu}

\cortext[cor1]{Corresponding author}

\address[isu]{Department of Materials Science and Engineering, 
Iowa State University, Ames, Iowa 50011}
% \address[deere]{John Deere, Materials Engineering Metals and Mechanics, Ag and Turf Division, MTIC One John Deere Place, Moline Illinois 61265 USA.}

\begin{abstract}
A new phenomenological technique for using constant amplitude loading data to 
predict fatigue life from a variable amplitude strain history is presented. 
A critical feature of this reversal-by-reversal model is that the 
damage accumulation is inherently non-linear. 
The damage for a reversal in the variable
amplitude loading history is predicted by approximating that the 
accumulated damage comes from a constant amplitude loading 
that has the strain range of the particular variable amplitude reversal. 
A key feature of this approach is that overloads at the beginning 
of the strain history have a more substantial impact on
the total lifetime than overloads applied toward the end of the cycle life.  
This technique effectively incorporates the strain 
history in the damage prediction and has the advantage 
over other methods in that there are no fitting parameters
that require substantial experimental data.  
The model presented here is validated using experimental variable 
amplitude fatigue data for three different metals. 
\end{abstract}


\end{frontmatter}

\section{Introduction\label{intro}}

One of the most prevalent methods for testing the fatigue properties 
of a material is to construct a constant amplitude strain-life curve.  In 
this method, a sample is strained cyclically between two strain levels 
until failure.  This test is performed on a number of identical samples 
at different strain magnitudes, and the results are plotted as a strain-life 
curve.  This ``$\frac{\Delta \varepsilon }{2}-2N_{f}$'' curve can be 
fitted to the Basquin-Manson-Coffin, \cite{basquin,manson,coffin} 
(BMC) equation, 
\begin{equation}\label{BMC}
\frac{\Delta \varepsilon }{2}=\frac{\Delta \varepsilon_e }{2}+\frac{\Delta \varepsilon_p }{2}=\frac{\sigma_{f}^{'}}{E}\left ( 2N_f \right )^b + \varepsilon_{f}^{'}\left ( 2N_f \right ) ^c ,
\end{equation}
where $\frac{\Delta \varepsilon}{2}$ is half of the strain range,  
$\Delta \varepsilon_e$  is the elastic strain, $\Delta \varepsilon_p$ 
is the plastic strain, $\sigma_{f}^{'}$ is the fatigue strength 
coefficient, $E$ is the elastic modulus, $b$ is the fatigue strength 
exponent, $\varepsilon_{f}^{'}$ is the fatigue ductility 
coefficient, $c$ is the fatigue ductility exponent, and  $2N_f$ is 
the number of reversals until failure.  The low cycle fatigue regime 
is considered a product of plastic deformation, and the high cycle 
portion is related primarily to elastic deformation. This approach 
gives a reasonably accurate prediction of sample life at constant 
amplitude cyclic strains.  

However, parts in service are rarely subjected to idealized 
constant amplitude cyclic strains and instead undergo variable 
amplitude loading.  The Palmgren-Miner (PM) rule is a widely 
used approach for predicting part lifetime under variable amplitude 
loading.  This method hypothesizes that the damage caused by 
each stress state in a variable amplitude load history is a function 
of the number of times that the particular cyclic stress state 
occurs and the number of cycles it would take for the sample 
to fail from a constant amplitude history at that stress state.~\cite{Miner}  
Mathematically the Palmgren-Miner rule is written, 
\begin{equation}\label{miner}
\sum_{i=1} \frac{n_i}{N_{fi}}=C,
\end{equation}
where the sum is taken over all stress states with $i$ denoting a 
particular state,  $n_i$ is the number of cycles at the $i$th stress 
state, $N_{fi}$ is the number of cycles to failure if the sample is 
cycled under constant amplitude loading at the $i$th stress state, 
and $C$ is a constant. Based on the assumptions made by the model, $C$ 
should be $1$; experimentally it is found to range between $0.7$ 
and $2.2$. This variability is evidence of the failure of the 
Palmgren-Miner rule to accurately predict fatigue lifetimes.  
Although it is well known that the Palmgren-Miner rule is 
inaccurate, its conceptual simplicity and the minimal amount of data 
necessary for implementation makes it a popular method 
for estimating fatigue life.  

Improving upon the Palmgren-Miner rule has been a major 
focus of researchers studying variable amplitude fatigue. 
A good review of cumulative damage and life prediction 
theories through the end of the last century is presented by Fatemi and Yang.  \cite{review1997}.
%Fatemi and Yang in reference \cite{review1997}. 
In recent years models have been 
developed based on a variety of techniques that couple theories of fracture 
mechanics and empirical 
observations.~\cite{Sun, Shang, ghammouri, Risitano, Huang, Besel, Shi,  Aid, Rejovitzky, tensiontorsionrandom}
Other methods take into account the residual stresses caused by the plasticity 
of the material at the crack tip, and crack tip closing 
phenomena.~\cite{Noroozi, Halliday, shanyavskiy}  Although these modeling methods 
are more accurate than the Palmgren-Miner rule, they require substantially 
more experimental data to fit the necessary parameters.

In this paper a new method, free of fitting parameters, is demonstrated for 
estimating strain life under variable amplitude 
loading.  
It is unique from other models in that the 
only data used for input is the constant 
amplitude strain-life curve and the cyclic 
stress-strain response.  
Using this model it is possible to accurately 
predict the variable amplitude strain life of specimens 
using a relatively small amount of experimental 
data that can easily be generated.  
In the section following this introduction the 
analytical details of the model are presented.  
Next the experimental procedure and 
model implementation are explained.  
In the results and discussion section the 
measured and predicted strain lifetimes 
are presented and compared.  The theoretical 
lifetimes predicted from this model are compared to lifetime predictions 
presented in the literature.
In the final section the paper is succinctly summarized.  

\section{Model Details\label{model}}

In this model the total damage, $D_T$, is the sum of the damage of all reversals, $D_i$, ranging from $i=1$ to $i = 2N_T$,
\begin{equation}\label{DTDEF}
D_T=\sum_{i=1}^{2N_T}D_i.
\end{equation}
In this definition $D_i$ is the normalized damage caused by the $i$th reversal and failure occurs when $D_T=1$. 
The damage caused by each reversal is determined using the well-known constant amplitude strain-life relation.  Using this approach, the strain history is incorporated as damage accumulation. The damage accrued in each step is calculated using a relatively simple algorithm and constant amplitude strain-life data.

Following examples from Ref.~\cite{review1997}, the damage during fatigue is assumed to be due to a single critical crack propagating across the width of the specimen.  A good description of the crack growth per reversal, for a constant amplitude strain, is given by the hyperbolic sine function.  For a crack size, $a$, the rate of crack tip advance, after $2N_k$ reversals, can be written 
\begin{equation}\label{dadN}
\frac{da}{dN} \propto \sinh \left (\frac{2N_k}{2N_f}\rho \right ),
\end{equation}
where $2N_f$ is the total number of reversals to failure for the given strain amplitude, and $\rho$ is a scaling factor, which will be discussed in more detail later.  Eq.~(\ref{dadN}) is a natural expression of damage per reversal, and has the same functional form as has been used to describe changes in crack growth rates previously.~\cite{frenchbook,review1997}   It accurately reflects the phenomena in that during the initial stages of damage the rate of crack propagation is low and as damage is accrued the rate of crack growth increases.  
Other expressions for the rate of crack tip advance, which are phenomenologically similar, are presented elsewhere and will not be discussed here.~\cite{frenchbook,review1997}

Using Eq.~(\ref{dadN}) for damage, the normalized damage due to the $i$th reversal is expressed, 
\begin{equation}\label{Di}\begin{aligned} 
D_i&=\frac{\sum_{k=1}^{N_T+1} \sinh \left (\frac{2N_k}{2N_f}\rho \right ) - \sum_{k=1}^{N_T} \sinh \left (\frac{2N_k}{2N_f}\rho \right )}{\sum_{j=1}^{N_f} \sinh \left (\frac{2N_j}{2N_f}\rho \right )}  \\ 
&\approx \frac{\int_{N_T}^{N_T+1} \sinh\left( \frac{2N_k}{2N_f}\rho \right )dN_k}{\int_{1}^{N_f} \sinh\left( \frac{2N_j}{2N_f}\rho \right )dN_j},
\end{aligned}
\end{equation}
where $N_T$ is the number of reversals required to achieve the 
accumulated damage, $D_T$, were the damage due to a constant amplitude strain range, $\frac{\Delta \varepsilon_i}{2}$. 
The number of reversals to failure at this constant strain range is  $2N_f$.  The denominator normalizes the damage per reversal such that at failure the total damage is 1.  

The current state of damage, $D_T$, is known and is expressed as the sum of incremental damage, $D_i$, from Eq.~(\ref{Di}),
\begin{equation}\label{DT} \begin{aligned}
D_T& = \frac{\sum_{k=1}^{N_T} \sinh \left (\frac{2N_k}{2N_f}\rho \right ) }{\sum_{j=1}^{N_f} \sinh \left (\frac{2N_j}{2N_f}\rho \right )} \\
&\approx  \frac{\int_{1}^{N_T} \sinh\left( \frac{2N_k}{2N_f}\rho \right )dN_k}{\int_{1}^{N_f} \sinh\left( \frac{2N_j}{2N_f}\rho \right )dN_j},
\end{aligned}
\end{equation}
where $N_T$ is the number of damaging reversals to cause the damage $D_T$ at the constant amplitude strain $\frac{\Delta \varepsilon}{2}$.  
The summations in Eqs.~(\ref{Di}) and~(\ref{DT}) can be approximated as integrals, which allows Eq.~(\ref{DT}) to be solved to find 
\begin{equation}\label{NT}\begin{aligned}
&2N_T= \left ( \frac{2N_f}{\rho} \right ) \times \\
&\cosh^{-1} \left ( D_T \left ( \cosh \left ( \rho \right ) - \cosh \left ( \frac{\rho}{2N_f} \right ) \right) + \right. \\
&\left. \cosh \left ( \frac{\rho}{2N_f} \right ) \right ).
\end{aligned}
\end{equation}

The general approach for calculating a specimen life 
under variable amplitude loading goes as follows.  
Begin by determining the initial total damage. %D_{T}$.  
This is typically near zero, if the part begins pristine and the strain-life curve is well defined at all strains of interest. 
A case for non-zero starting damage will be discussed later in section \ref{modelimplementation}.
The strain range for the first tensile reversal is  $\frac{\Delta \varepsilon_1}{2}$, and $D_1$ is calculated using 
Eq.~(\ref{Di}) and the BMC relation, Eq.~(\ref{BMC}), which is fit to constant amplitude strain life data.  
The value of $D_1$ is added to $D_T$.  The next tensile strain is $\frac{\Delta \varepsilon_2}{2}$.  
The lifetime at this strain range along with $D_T$, are used to calculate $N_T$ 
from Eq.~(\ref{NT}), which is used in Eq.~(\ref{Di}) to calculate $D_2$, which is again 
added to $D_T$. This process continues until $D_T=1$, at which point failure is predicted.
An algorithm for implementing this model is demonstrated in section \ref{modelimplementation}.

\section{Experimental Data\label{experiment}}

Published experimental data are used to validate the model.  \cite{Pereira,Colin}
 Pereira et  al.~tested P355NL1 steel and compared their experimental results to an effective strain damage model based 
on the work of DuQuesnay. \cite{Pereira}
Colin and Fatemi published experimental data for 304L stainless steel and 7075 T6 aluminum.  \cite{Colin} 

\begin{figure}[h!]
		\includegraphics[scale = 0.35]{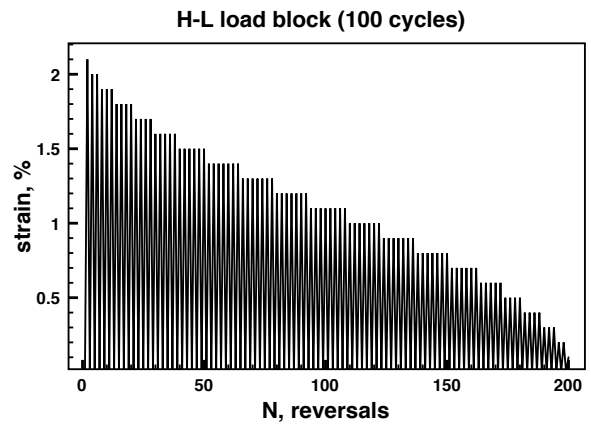}
		\caption{An example of a high to low strain loading
                  block after the work of Pereira et al.~\cite{Pereira} }
		\label{fig1}
	\end{figure}

\begin{figure}[h!]
		\includegraphics[scale = 0.35]{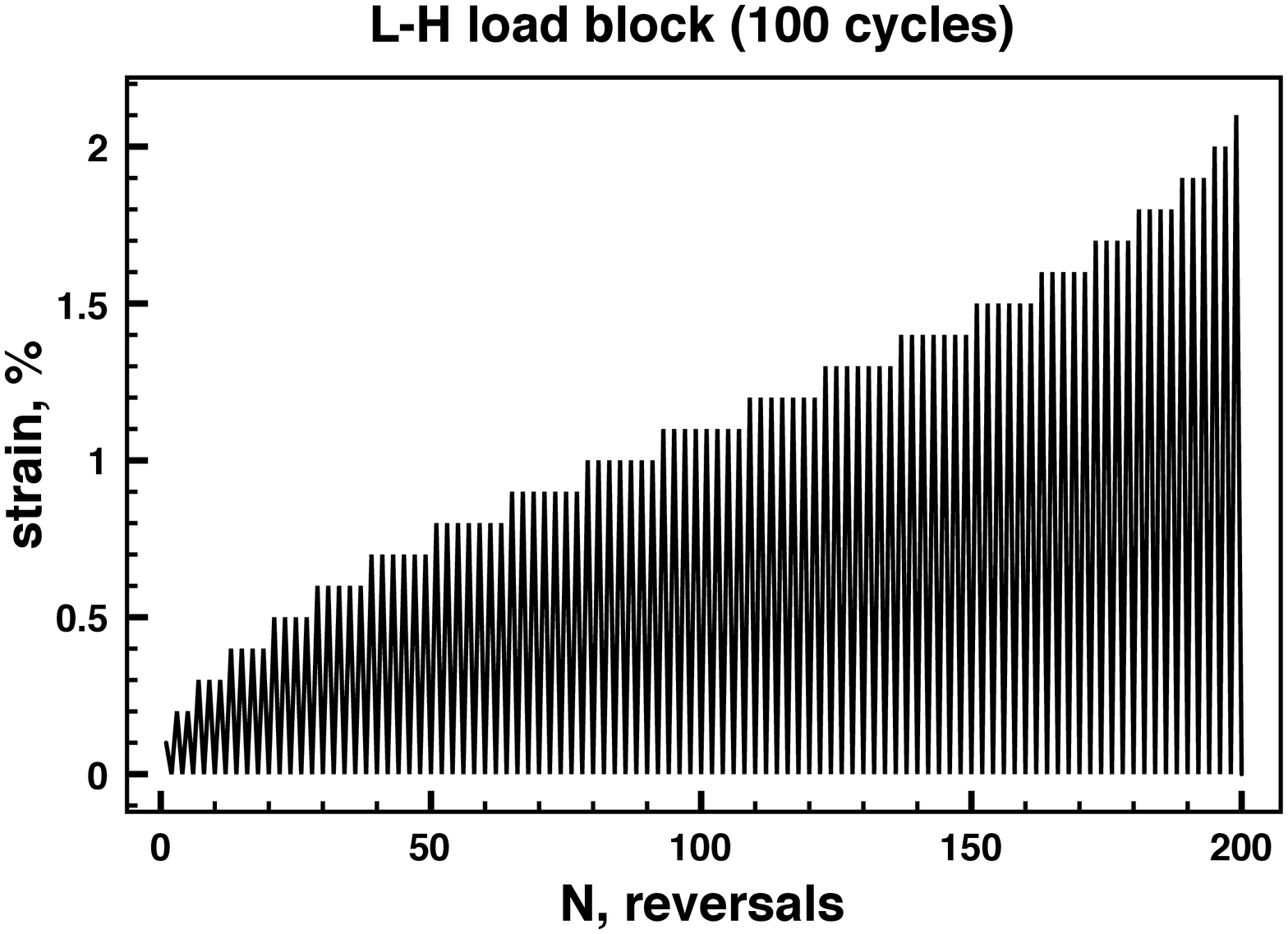}
		\caption{An example of a low to high strain loading block after the work of Pereira et al.~\cite{Pereira} }
		\label{fig2}
	\end{figure}

\begin{figure}[h!]
		\includegraphics[scale = 0.35]{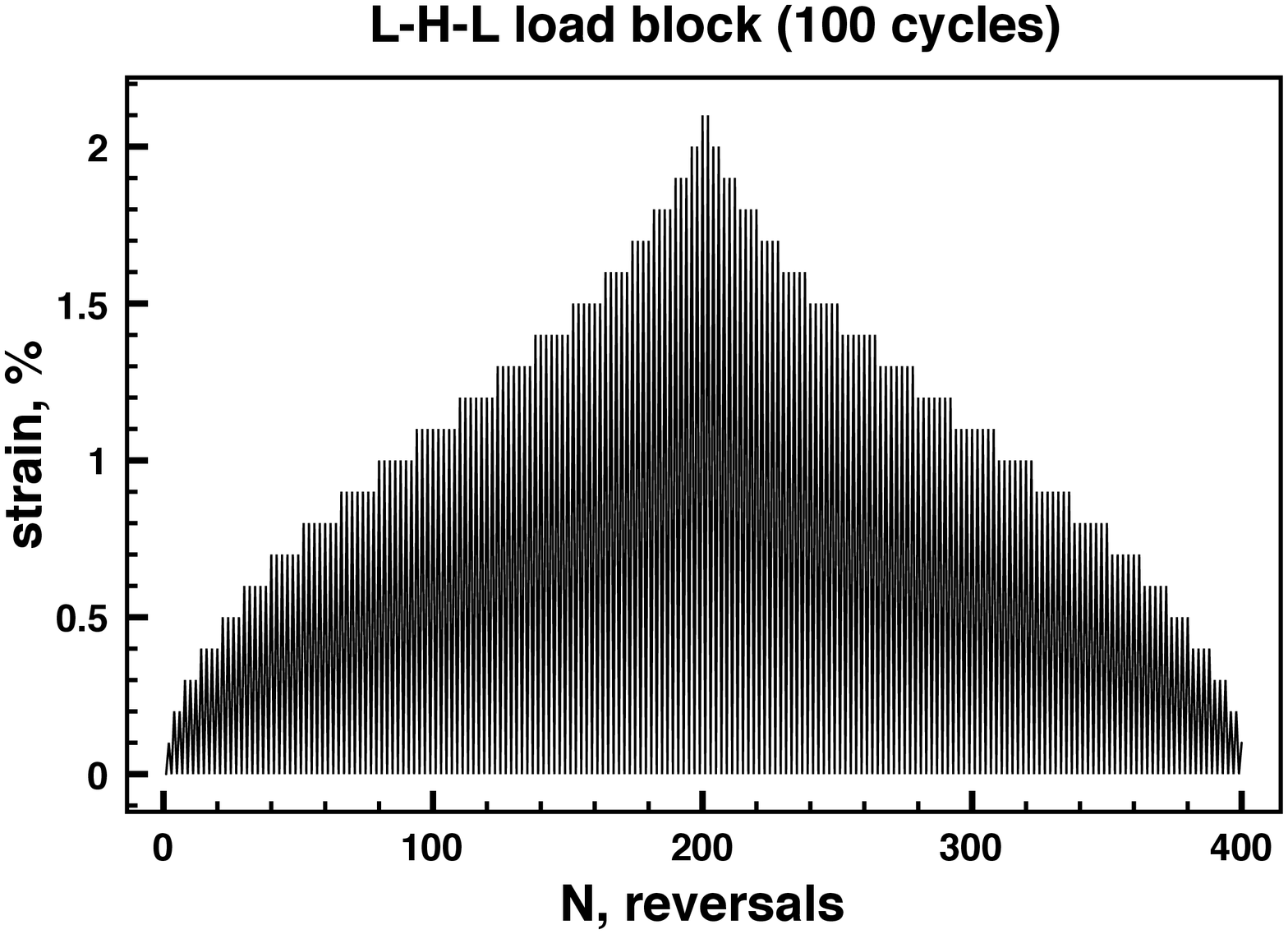}
		\caption{An example of a low to high to low strain loading block after the work of Pereira et al.~\cite{Pereira} }
		\label{fig3}
	\end{figure}

\begin{figure}[h!]
		\includegraphics[scale = 0.35]{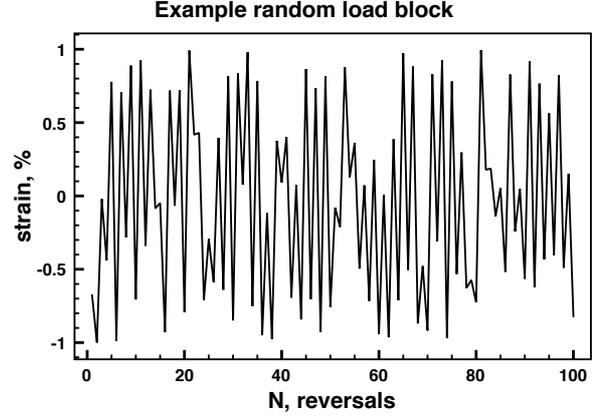}
		\caption{An example of random loading. Note that the
                  random number stream was filtered to 
ensure that each iteration reversed the direction of the applied strain.}
		\label{fig4}
	\end{figure}

\begin{figure}[h!]
		\includegraphics[scale = 0.35]{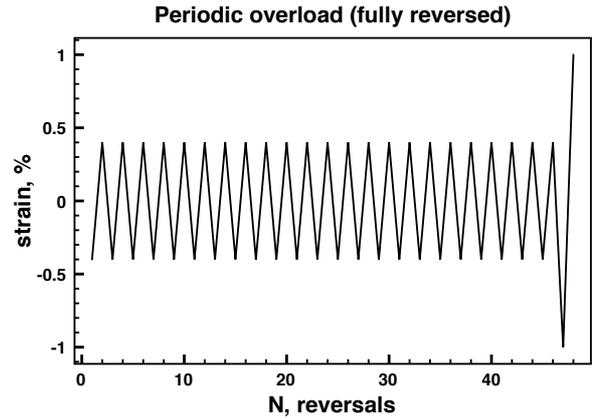}
		\caption{An example of a fully reversed periodic overload loading block after the work of Colin and Fatemi.~\cite{Colin}}
		\label{fig5}
	\end{figure}

The load histories used in the calculations were recreated from the descriptions detailed by Pereira et  al., and
Colin and Fatemi.  \cite{Pereira,Colin}
The P355NL1 specimens were subjected to a variety of load blocks,
including a high to low scheme, a low to high scheme, 
a low to high to low scheme, and random loading, examples of which are
shown in Figs.~\ref{fig1}, \ref{fig2}, \ref{fig3}, and \ref{fig4}.
The 304L stainless steel fatigue-life 
results were prepared for periodic, fully-reversed overloads, shown in Fig.~\ref{fig5}, and random loading.  
The 7075 T6 aluminum samples were subjected to random loading.  
Unlike the shaped loading blocks, the random load history was not reproduced from the literature, but 
instead a strain history file was created using a random number
generator, filtering the random number 
stream to ensure that each iteration reversed the strain. A
representative sample of the random reversal data 
is shown in Fig.~\ref{fig4}.

%\begin{figure}[h!]
	
%	\caption{}
%	\label{loads}
%\end{figure}

\section{Model Implementation\label{modelimplementation}}

The model described in section \ref{model} above was implemented using the algorithm 
detailed in this section and applied to the strain history data discussed in section \ref{experiment}.
To begin, the first two strains from the strain history were 
converted to stresses using the 
Ramberg-Osgood stress-strain relationship,
 \begin{equation}\label{rambergosgood}
 \varepsilon = \frac{\sigma}{E}+\left ( \frac{\sigma}{K^{'}} \right ) ^{1/n'},
 \end{equation}
 where $K^{'}$ and $n'$ were fit from the cyclic stress-strain behavior.  
The Morrow mean stress correction \cite{morrow},
\begin{equation}\label{stresscorrection}
\frac{\Delta \varepsilon}{2}=\frac{\sigma^{'}_f-\sigma_m}{E} \left (2N_f \right )^{b} + \varepsilon^{'}_f \left ( 2N_f \right )^{c},
\end{equation}
where $\sigma_m=\frac{\sigma_{i+1}+\sigma_i}{2}$ is the mean stress, was used to 
calculate $N_{f}$. 
The value of $N_{f}$ was used with $D_{T}$ to 
calculate $N_T$ from Eq.~(\ref{NT}).  
Finally, the damage from this reversal, $D_i$, was determined from Eq.~(\ref{Di}). 
The damage was added to the total damage, $D_T$.  If the total damage was greater 
than $1$, then the specimen was deemed to have failed due to this reversal, 
otherwise the procedure was continued using the new value of $D_T$ and the 
next strain taken from the strain history.  This was repeated until failure occurs, when $D_T\ge1$.

The constant $\rho$, in Eq.~(\ref{dadN}), appropriately 
scales the incremental damage. 
For this implementation it was selected to 
be $\rho = \log \left ( 2N_f^\gamma \right )$, where $2N_{f}$ was 
from the BMC equation and $\gamma$ was taken as $-c/2\varepsilon_f^{'}$.
In this way, $\rho$ scales with the applied strains and the incremental 
damage, $da/dN$, has the correct functional relationship to the strain amplitude.  

The integrated $da/dN$ curve, which expresses total damage as a function of the number of cycles,
has a general shape that is well known.~\cite{review1997,frenchbook,Sun,Shang}
Careful inspection of experimental data of damage as a function of number of cycles reveals that 
a smooth well fitting curve does not always intercept the damage axis at $0$.
In many experiments, damage has been observed to accumulate rapidly to around 5 to 10\% early in the specimen's life, and then 
slow to the $da/dN$ crack propagation model that is well known.~\cite{frenchbook,Sun,Shang}
To account for the rapid damage accumulation that occurs during the initial cycling, 
the starting damage was assigned to be $D_T=0.05$ for all of the variable amplitude lifetime prediction data 
presented here.  

\section{Results and Discussion}\label{results}
\begin{figure}[h!]
		\includegraphics[scale = 0.321]{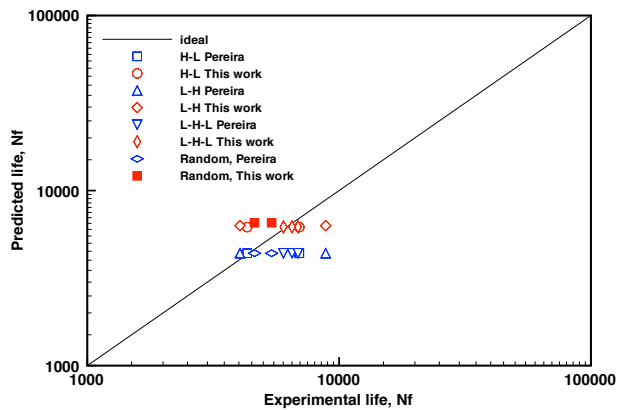}
		\caption{A comparison of this work and the work of
                  Pereira et al.~for P355NL1 
steel specimens tested by Pereira et al.~and presented in Ref.~\cite{Pereira}.  
A variety of load schemes are used at a maximum strain of 1.05\%.  }
		\label{fig6}
	\end{figure}

\begin{figure}[h!]
		\includegraphics[scale = 0.33]{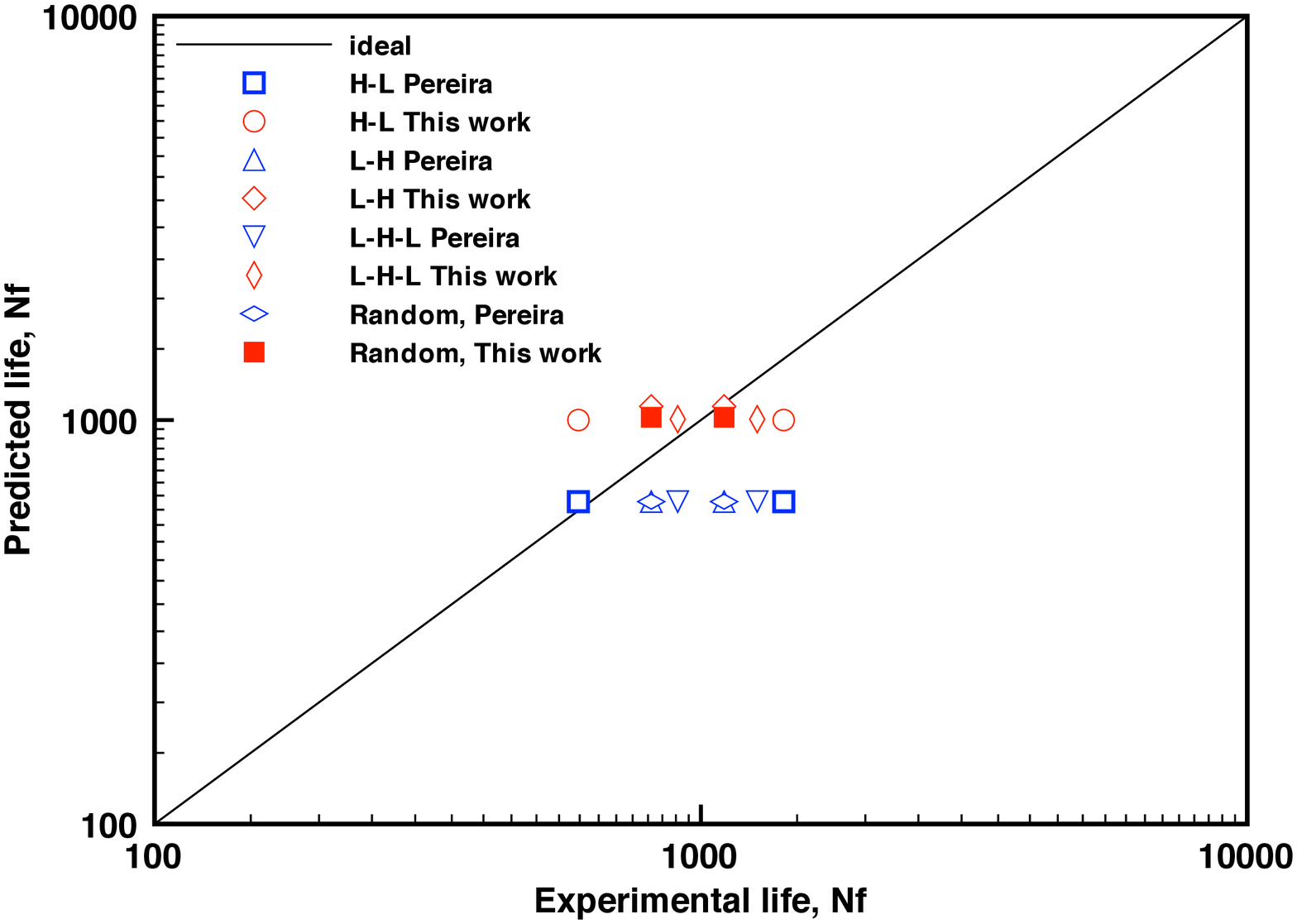}
		\caption{A comparison of this work and the work of 
Pereira et al.~for P355NL1 
steel specimens tested by Pereira et al.~and presented in Ref.~\cite{Pereira}.
A variety of load schemes are used at a maximum strain of 2.1\%. }
		\label{fig7}
	\end{figure}

\begin{figure}[h!]
		\includegraphics[scale = 0.282]{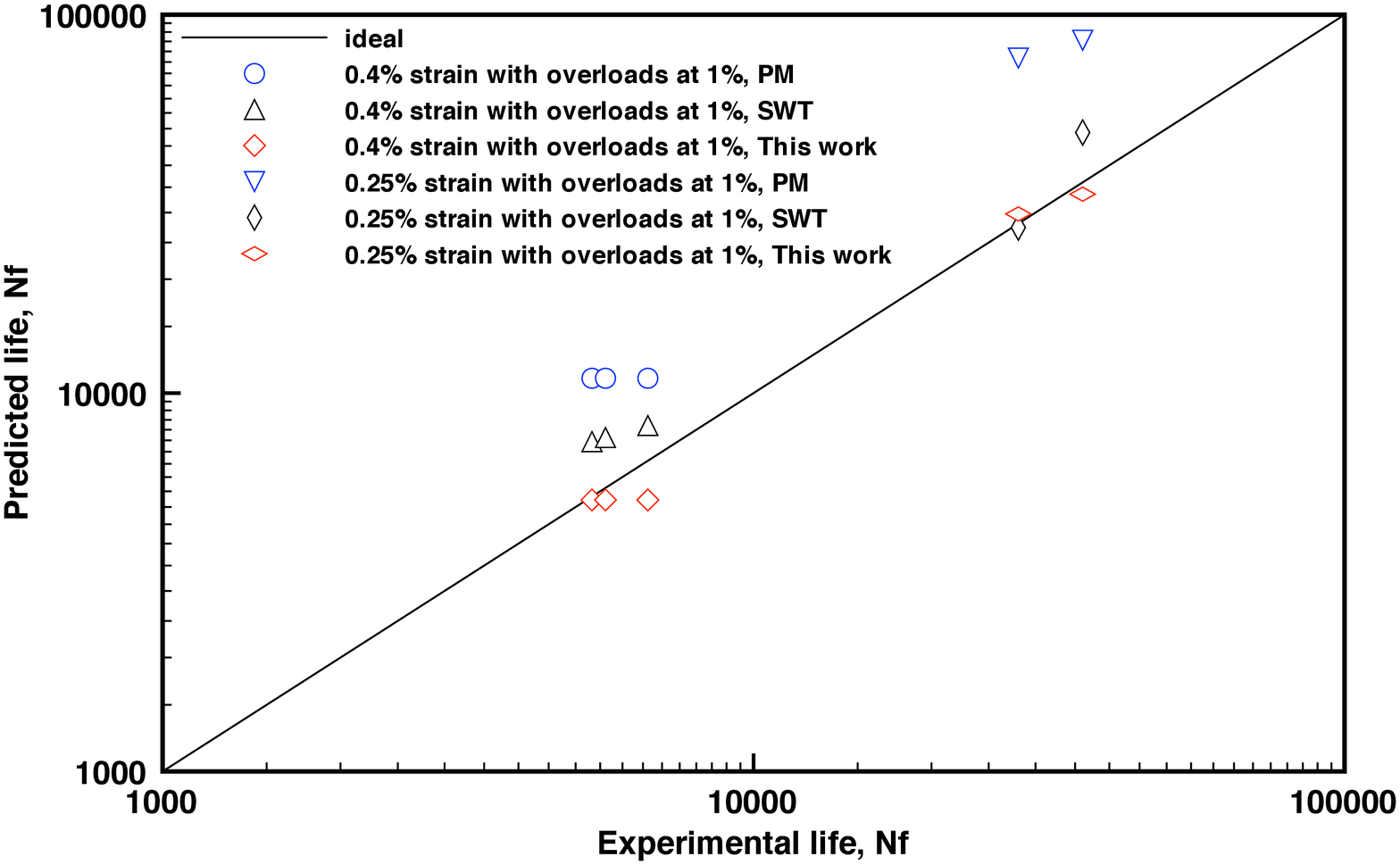}
		\caption{A comparison of the predictions of this work to the Palmgren-Miner rule (PM) with and without a Smith-Watson-Topper correction (SWT) prepared by Colin and Fatemi for 304L stainless steel specimens.~\cite{Colin}
The data here represent specimens subjected to a constant amplitude fully reversed strain history punctuated by periodic fully reversed overloads. }
		\label{fig8}
	\end{figure}

\begin{figure}[h!]
		\includegraphics[scale = 0.252]{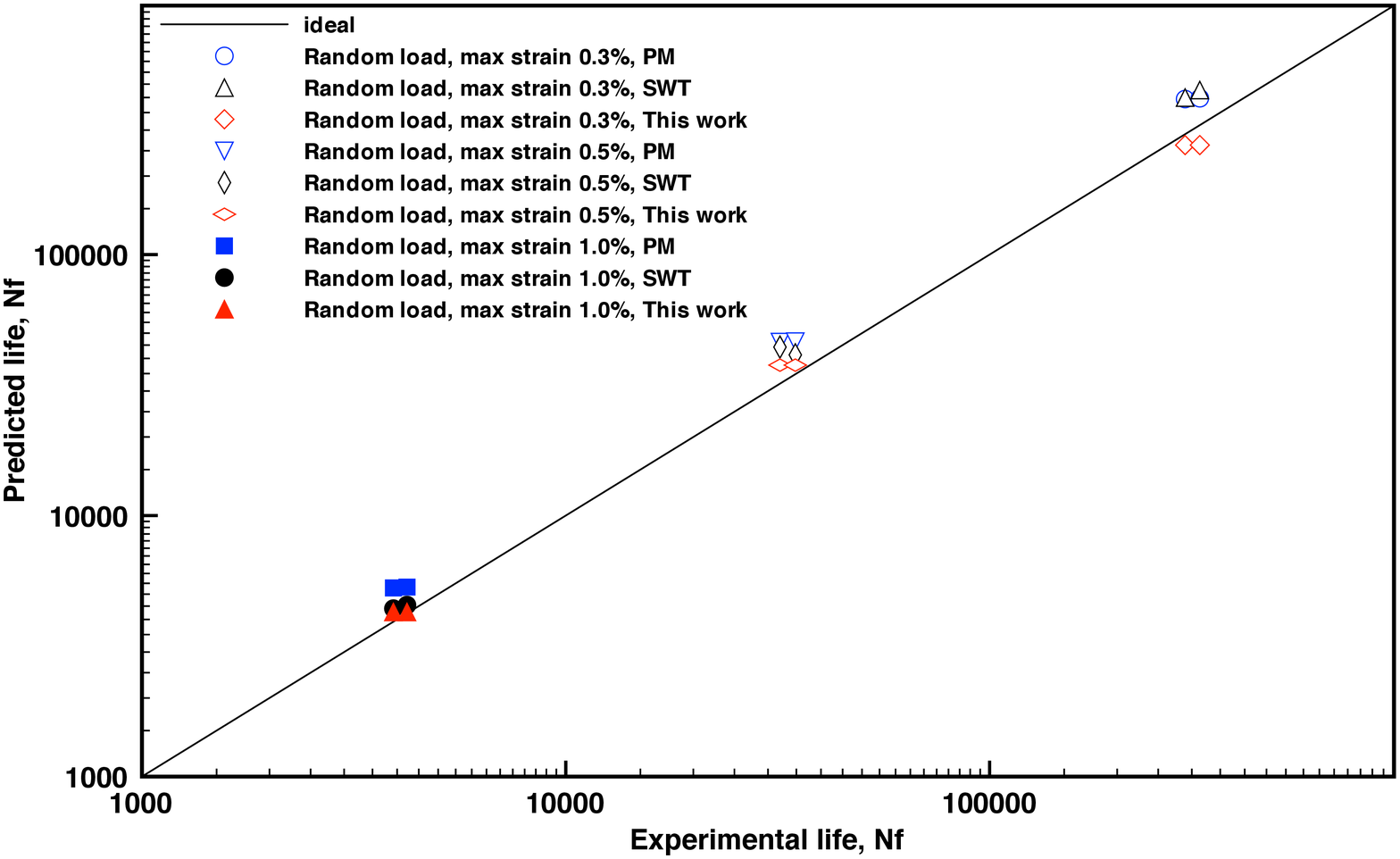}
		\caption{A comparison of the predictions of this work to the Palmgren-Miner rule (PM) with and without a Smith-Watson-Topper correction (SWT) prepared by Colin and Fatemi for 304L stainless steel specimens.~\cite{Colin}The data
here represent specimens subjected to random loading.}
		\label{fig9}
	\end{figure}

\begin{figure}[h!]
		\includegraphics[scale = 0.252]{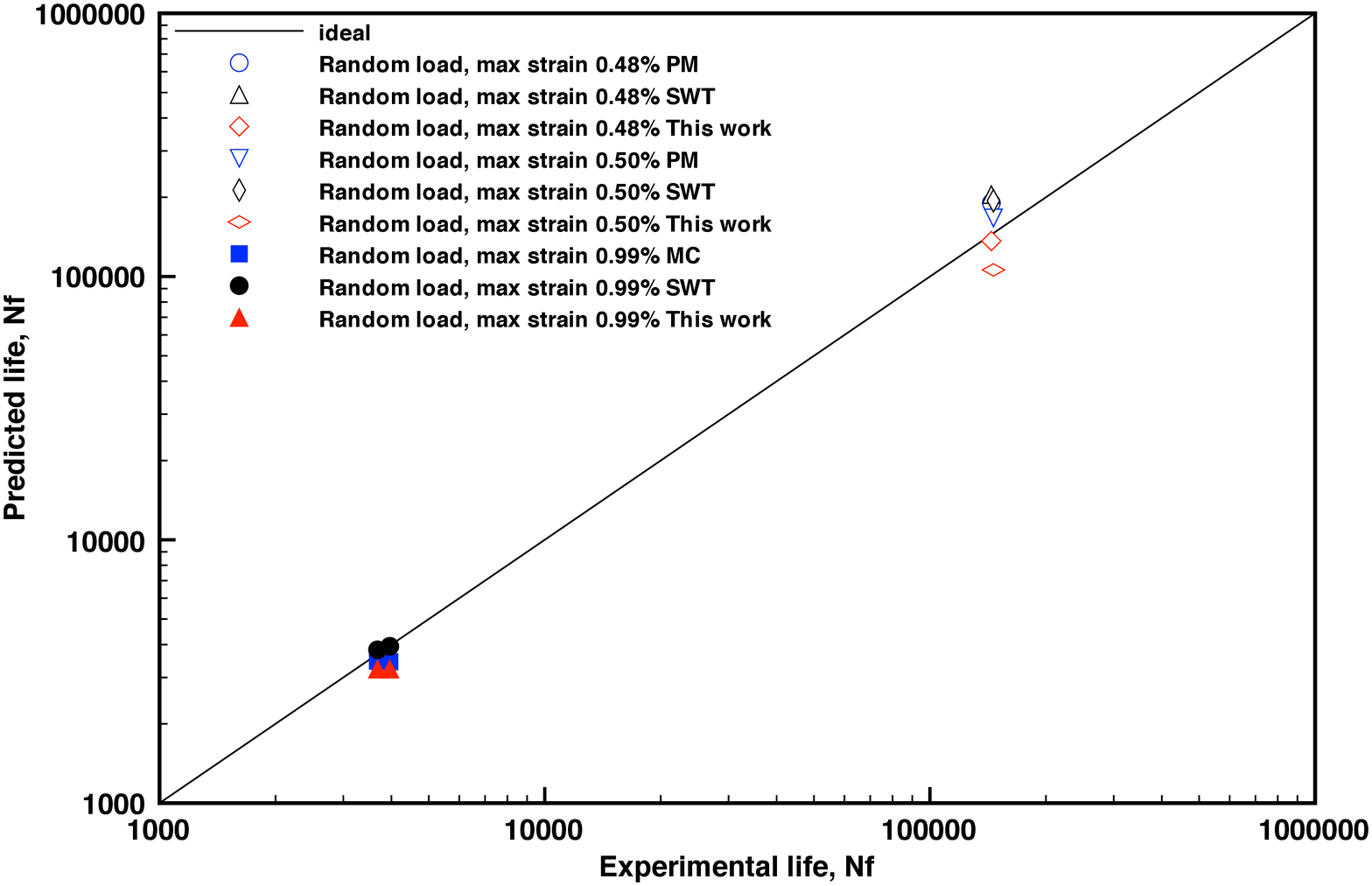}
		\caption{A comparison of the predictions of this work to the Palmgren-Miner rule (PM) with and without a Smith-Watson-Topper correction (SWT) prepared by Colin and Fatemi for 7075-T6 aluminum specimens.~\cite{Colin} The data
here represent specimens subjected to random loading.}
		\label{fig10}
	\end{figure}

A comparison of the published experimental data and the predictions 
of this work can be seen in Figs.~\ref{fig6}--\ref{fig10}.
In Figs.~\ref{fig6} and \ref{fig7} the experimental data for P355NL1
steel, from Pereira et  al., is 
shown for a maximum strain of 1.05\% and 2.10\% for the loading blocks discussed in 
section \ref{experiment} above.~\cite{Pereira} The experimental results are compared to those predicted from 
the present model in addition to the model of Pereira and DuQuesnay.  
The current model is in good agreement with both the experimental and theoretical results 
from Peirera et  al.

The experimental results from Colin and Fatemi for 304L stainless steel are shown in 
Figs.~\ref{fig8} and \ref{fig9} for the loading blocks discussed above at various strain ranges.~\cite{Colin}
The results for 7075 T6 aluminum are shown in Fig.~\ref{fig10}.  The accuracy of the present model 
is compared to the predictions of the Palmgren-Miner rule both with and without the 
Smith-Watson-Topper correction. The current model is again in good agreement with the 
experimental results. It is worth noting that the model presented here agrees with the experimental data both when 
the Palmgren-Miner results are non-conservative by an order of 2 or 3 and when they agree 
well with experiment.  This is evidence that the results form the current work are 
more profound than a simple lifetime reduction from the Palmgren-Miner rule.

%\begin{figure}[h!]

%	\caption{}
%	\label{results}
%\end{figure}

The strength of the present model is due to the natural inclusion of the 
strain history when determining the inflicted damage caused by a strain reversal.  
Both the effect of 
the immediately preceding strain and the effect of the relative
age of a specimen are included.  
To calculate $\Delta \varepsilon$ a strain and
the immediately preceding strain must be known to determine the applied strain 
range and the mean stress correction.  
The damage inflicted by a particular strain reversal depends not only on the amplitude, 
but also on the total state of damage at the instant of the reversal.  
Take the example of a periodic, fully-reversed, overload.  If one considers 
the integrated $da/dN$ curve for a constant amplitude cyclic strain,
applying a fully reversed overload 
would advance the position of the subsequent constant amplitude reversals on this curve substantially.  
From the shape of the curve, it is apparent that an overload early in the specimen's life that  
increments the damage will have a more substantial impact on the specimen's total life, 
compared to an overload applied later.

The more recent models, such as those of \cite{Noroozi, Halliday, shanyavskiy}, 
include residual strains and crack-tip plasticity or crack tip closure phenomena. 
These models account for the cumulative damage through localized plasticity 
near the crack tip, or a change in the effective stresses due to crack tip closure.  
They are more accurate than the Palmgren-Miner model and other simple 
models because they account for the ordering of the 
applied strains and localized damage near the crack tip; however, 
they require substantial experimental support.  
By comparison, the present model only includes materials information from the 
cyclic stress-strain curve and the constant amplitude strain life curve.  
%Although this simplified approach does not carry with it the detailed information 
%inherent in the more sophisticated models, it represents a substantial improvement over the 
%Palmgren-Miner model and other simple models while requiring relatively little experimental 
%input.  
It should be noted that for some analysis involving extreme load cases, a more 
sophisticated model than the one presented here may be necessary.  In particular, cases
where residual plasticity will play a dominant role in crack growth rates through strain
hardening and residual stresses will likely require a model that explicitly deals with plasticity
near the crack tip.

\section{Summary}

One of the greatest engineering challenges 
of the last $100$ years is that of predicting the 
strain-life relationships of mechanical components 
undergoing variable amplitude loading.  
In spite of extensive studies,
no conclusive model has been determined.  
Although many useful models
have been developed, many require cumbersome amounts of experimental data.
Here we report a new method, free of fitting parameters, for making an accurate variable 
amplitude strain-life prediction using basic 
constant amplitude fatigue data.  
The method is 
validated using data from experiments performed on P355NL1 steel, 
304L stainless steel, and 7075 T6 aluminum.~\cite{Pereira,Colin}
The present model fits the experimental data 
well for a variety of load spectra and materials and the algorithm is simple 
to implement. 

\section{Acknowledgements}

The authors gratefully acknowledge funding from John Deere \& Company.  

%\bibliographystyle{model1-num-names}
%\bibliography{fatigue}

\end{document}